\documentclass[10pt,draft]{dis03}
\usepackage{epsf,amsmath}

\newcommand{\beq} {\begin{equation}}
\newcommand{\eeq} {\end{equation}}
\newcommand{\beqa} {\begin{eqnarray}}
\newcommand{\eeqa} {\end{eqnarray}}

\newcommand{\jpsi}{J/\psi}

\newcommand{\lsim}{\buildrel < \over {_\sim}}

\begin{document}

\title{
\hbox to\hsize{\normalsize\hfil\rm LAPTH-Conf-990/03}
\hbox to\hsize{\normalsize\hfil hep-ph/0307304}
\hbox to\hsize{\normalsize\hfil July 24, 2003}
\vskip 30pt
\centerline{COMOVER ENHANCEMENT SCENARIO}
\centerline{FOR QUARKONIUM PRODUCTION\footnote{To appear in the proceedings
of the XIth International Workshop on Deep Inelastic Scattering,  
St. Petersburg, 23-27 April 2003.}}} 
\author{S.~Peign\'e \\
LAPTH, B.P. 110, F-74941 Annecy-le-Vieux Cedex, France\\
E-mail: peigne@lapp.in2p3.fr}
\maketitle

\begin{abstract}
\noindent The quarkonium data suggest a qualitatively new production 
mechanism, enhancing the rate of the quarkonia produced  
in the fragmentation region of an incoming coloured particle.
\end{abstract}

\section{Introduction and summary}  
The production of heavy $Q\bar Q$ quarkonia is a promising probe of 
QCD dynamics, both in its perturbative and non-perturbative aspects.
However, in quarkonium production even a qualitative agreement 
between theory and experiment is not yet achieved.
In such a situation, it might be useful to recall the
qualitative features of the quarkonium data, in order to 
gain some possibly new insight.

In this talk I argue that the main features of the 
quarkonium production data suggest a new mechanism for 
quarkonium low $p_{\perp}$ hadroproduction, which breaks
factorization between the perturbative production of the heavy 
$Q\bar Q$ pair and its subsequent hadronization into a 
bound state. The mechanism is based on the rescattering of the
$Q \bar Q$ pair on the comoving field created in $gg$ fusion, and  
appears to be qualitatively consistent with many
features of the quarkonium production data.  
We expect a similar mechanism to occur each time the $Q \bar Q$ pair
is produced in the fragmentation region of a coloured 
particle, and thus surrounded by the DGLAP field radiated by this
particle. 

\section{The qualitative difference between hadro and photoproduction}
The first well-known fact I would like to recall is the failure of the
Color Singlet Model (CSM) in predicting quarkonium
{\it hadroproduction} rates: $\jpsi$ and $\psi'$ hadroproduction
rates are underestimated by more than an order of magnitude,
both at low $p_{\perp} \lsim m_c$ \cite{Kaplan:1996tb} 
and large $p_{\perp} \gg m_c$ (the $\psi'$ anomaly) \cite{Kramer:2001hh}. 
We stress however that at large $p_{\perp}$, the {\it shape} of 
$d\sigma(p{\bar p} \rightarrow \psi +X)/dp_{\perp}$ is correctly given
by the perturbative part of the process, which is dominated by gluon 
fragmentation \cite{Braaten:1993rw}. Even though the fragmentation
channel appears beyond leading-order (LO) in the CSM, 
{\it the CSM predicts correctly the shape of $d\sigma/dp_{\perp}$},
as does in fact any model where the $Q\bar Q$ pair is produced 
perturbatively.

In contradistinction with the clear failure of the CSM in 
$\psi$ hadroproduction, the CSM works surprisingly well when the
projectile hadron is replaced by a photon. Namely, 
{\it the CSM is consistent with quarkonium photoproduction}.
\begin{itemize}
\item{} In inelatic photoproduction, the CSM at NLO agrees well with
  the data, up to $p_{\perp} \simeq 7\ {\rm GeV} > m_c$ and for
  values of $z>0.3$\footnote{The $\jpsi$ rate thus arises
  from the photon fragmentation region, and resolved photon processes
  are negligible here.} \cite{H1,ZEUS}. We note that the CSM LO process
  $\gamma g \rightarrow \jpsi g$ is clearly insufficient, since the dominant
  photoproduction channel at $p_{\perp} > m_c$, namely 
$\gamma g \rightarrow \jpsi gg$, appears at NLO. 
\item{} In DIS, the $\jpsi$ production data from H1 have been compared to
  the CSM at LO \cite{Kniehl:2002jg}. The LO CSM underestimates the
  DIS $\jpsi$ rate by a factor $K \sim 2-3$ (to be contrasted with 
  $K > 10$ in hadroproduction), and does not yield the correct
  $p_{\perp}$-shape. The above reasonable `$K$-factor' is likely to be
  explained by NLO contributions and, again, CSM at LO is insufficient
  since it misses the dominant channel at $p_{\perp} \gg m_c$. It is clear
  that NLO contributions will improve the CSM prediction, both in
  shape and normalization. 
\item{} In $\gamma \gamma$ collisions, the CSM has been recently
  shown to underestimate the $\jpsi$ production rate (at 
  $p_{\perp} \leq 3\  \rm{GeV}$) by one order of magnitude
  \cite{Klasen:2002vc}. This may not be surprising, since the
  $\gamma \gamma \rightarrow \jpsi + X$ rate is actually do\-mi\-na\-ted
  by single resolved photon processes \cite{Klasen:2002vc}, namely
  $\gamma g \rightarrow \jpsi + X$, and the measured rate is
  integrated over a $\jpsi$ rapidity $-2<y_{\jpsi}<2$. The
  failure of the CSM in the present case might be due to an
  underestimation of the rate in the resolved photon fragmentation
  region, in which case it would be directly expected from the 
  similar failure in low $p_{\perp}$ hadroproduction.
\end{itemize}

\section{A new mechanism for quarkonium hadroproduction}
We know that the CSM is not a consistent model for quarkonium
production. In particular, contributions from the Color Octet Model
(COM) are necessary to make inclusive $P$-wave decay rates infrared
finite. On the other hand, the COM has several difficulties when
confronted to the data. The COM prediction for $\psi$ polarization
at the Tevatron seems to disagree with the observation
\cite{Braaten:1999qk}, and the COM tends to overestimate 
quarkonium photoproduction rates \cite{Kramer:2001hh}.

Our attitude is to suppose that the success of the CSM in
photoproduction is not fortuitous, and suggests the presence 
of a new production mechanism, occurring in hadroproduction, 
and more generally when the $Q \bar Q$ pair is produced 
in the fragmentation region of a coloured particle 
participating to the hard $Q \bar Q$ creation process.  
\begin{figure}[h]
\vspace*{3cm}
\begin{center}
\includegraphics{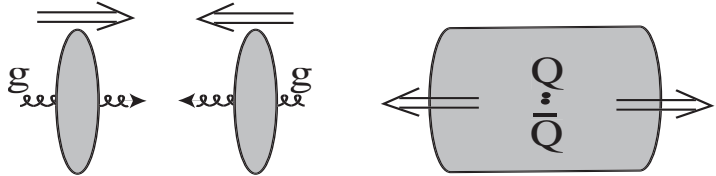}
\caption[*]{Quarkonium hadroproduction induced by the rescattering of the
  heavy $Q\bar Q$ pair off the surrounding DGLAP-originated color field.}
\end{center}
\end{figure}

\vskip -5mm
At high energy, quarkonium hadroproduction is dominated by $gg$
fusion, as pictured in Fig.~1. 
Each of the incoming gluons carries its proper brems\-strah\-lung DGLAP
field, and the heavy $Q\bar Q$ pair is thus produced in a rich comoving 
environment, in any fragmentation region. 
In non-resolved photoproduction, this environment is absent in the
photon fragmentation region (but should be present in the target
hadron fragmentation region). Similarly, it will
also appear in $\gamma \gamma$ collisions, in the fragmentation region
of the incoming photon which is resolved.  

We have suggested that semi-hard rescatterings between the $Q\bar Q$
pair and this comoving field could explain the observed `anomalies' of 
quarkonium production \cite{hmp}.
Qualitatively, this `Comover Enhancement Scenario' (CES) 
allows to explain the discrepancy of
the CSM with quarkonium hadroproduction and resolved photoproduction
(or resolved $\gamma \gamma$ collisions). The scenario is consistent
with the success of the CSM in the photon fragmentation region of 
(non-resolved) photoproduction. 

In low $p_{\perp}$ hadroproduction, the comoving field is assumed to
arise from the DGLAP evolution of the incoming gluons (Fig.~1). 
Since (at least) one rescattering between this field and the $Q\bar Q$ 
pair is used to create the quantum numbers of the 
bound state finally produced \cite{hmp}, our scenario breaks the
factorization usually assumed between the $Q\bar Q$ pair production
process and its subsequent color neutralization. This
mechanism is thus not included in the COM expansion, 
although there is a priori no reason to neglect it. 

Large $p_{\perp} \gg m_Q$ quarkonium hadroproduction is dominated 
by the fragmentation of a quasi-real gluon, 
$g(p_{\perp}) \rightarrow Q\bar Q$ \cite{Braaten:1993rw}. 
Since interactions between the $Q \bar Q$ pair
and the DGLAP fields of the incoming partons are suppressed by powers
of $1/p_{\perp}$, it is tempting to generalize our low $p_{\perp}$
scenario to large $p_{\perp}$ production by considering the DGLAP
field of the fragmenting gluon itself as a source of comovers. 
Any possible further rescattering with the $Q \bar Q$ pair respects
factorization, but it is not clear whether this type of process 
is included in the COM. Little is known on the convergence of the COM
expansion, and we cannot exclude that high orders in $\alpha_s$ and
$v$ are phenomenologically important, as suggested by the failure of 
the COM at first orders in predicting $\psi$ polarization at large
$p_{\perp}$. 

We have shown that assuming the relevance of the 
Comover Enhancement Scenario (CES) and of
the color singlet quarkonium wave function allows to explain many
features of the quarkonium production data \cite{hmp}. 
We summarize below the main
successes of the CES and refer to Ref.~\cite{hmp} for more details.
\begin{itemize}
\item{} The CES leaves intact the success of the CSM in direct photon
  processes. 
\item{} Since the comoving field arises from DGLAP evolution, the CES
  is a leading-twist mechanism, consistent with the fact that
  the `anomalies' of quarkonium production subsist when $\sqrt{s}$,
  $p_{\perp}$, or $m_Q$ are increased. 
\item{} The ratios of different $S$-wave quarkonium hadroproduction rates are
  consistent with the ratios of the (squared) color singlet wave
  functions at the origin. 
\item{} We expect the CES to enhance hadroproduction 
  rates of $C=-1$ states and $\chi_{1}$, which require three
  hard gluons to be produced. In this respect the fact that the 
  CSM underestimates low $p_{\perp}$ $\chi_{c1}$ hadroproduction is
  not surprising. 
\item{} Qualitatively, the CSM is consistent with inelastic (and
  non-resolved) $\chi_{c1}$ and $\chi_{c2}$ photoproduction 
  \cite{Roudeau:1988wb}. This will not be affected by the CES.
\item{} Within the CES we expect quarkonium nuclear absorption in $pA$
  collisions to occur because of the absorption of the incoming DGLAP
  fields, resulting in a comoving field being attenuated
  compared to $pp$ collisions. The CES thus 
  suggests that nuclear absorption might be due to a {\it lack} of
  hard comovers, contrary to conventional explanations.  
\end{itemize}

\section*{Acknowledgements} 
This talk is based on a collaboration with P.~Hoyer and N.~Marchal.

\end{document}